\documentclass[11pt]{iopart}
\usepackage{iopams,graphicx,cite,color}  
\newcommand{\gdw}{g$_{{\rm DW}}^{-1}$}
\newcommand{\hh}{h$^{-1}$}

\begin{document}

\title[Growth against entropy in bacteria]{Growth against entropy in bacterial metabolism: the phenotypic trade-off behind empirical growth rate distributions in {\it E. coli}}

\author{Daniele De Martino$^{\star}$, Fabrizio Capuani$^\S$, Andrea De Martino$^{\dag,\P}$}

\address{$^\star$ Institute of Science and Technology Austria (IST Austria), Am Campus 1, Klosterneuburg A-3400, Austria\\
$\S$ Dipartimento di Fisica, Sapienza Universit\`a di Roma, Rome, Italy\\
$\dag$ Soft and Living Matter Lab, Institute of Nanotechnology (CNR-NANOTEC), Consiglio Nazionale delle Ricerche, Rome, Italy\\
$\P$ Human Genetics Foundation, Turin, Italy}
%\ead{submissions@iop.org}
\vspace{10pt}
%\begin{indented}
%\item[]
%\end{indented}

\begin{abstract}
The solution space of genome-scale models of cellular metabolism provides a map between physically viable flux configurations and cellular metabolic phenotypes described, at the most basic level, by the corresponding growth rates. By sampling the solution space of {\it {\it E. coli}}'s metabolic network, we show that empirical growth rate distributions recently obtained in experiments at single-cell resolution can be explained in terms of a trade-off between the higher fitness of fast-growing phenotypes and the higher entropy of slow-growing ones. Based on this, we propose a minimal model for the evolution of a large bacterial population that captures this trade-off. The scaling relationships observed in experiments encode, in such frameworks,  for the same distance from the maximum achievable growth rate, the same degree of growth rate maximization, and/or the same rate of phenotypic change. Being grounded on genome-scale metabolic network reconstructions, these results allow for multiple implications and extensions in spite of the underlying conceptual simplicity.  
\end{abstract}

% Uncomment for PACS numbers
%\pacs{00.00, 20.00, 42.10}
%
% Uncomment for keywords
%\vspace{2pc}
%\noindent{\it Keywords}: XXXXXX, YYYYYYYY, ZZZZZZZZZ
%
% Uncomment for Submitted to journal title message
%\submitto{\JPA}
%
% Uncomment if a separate title page is required
%\maketitle
% 
% For two-column output uncomment the next line and choose [10pt] rather than [12pt] in the \documentclass declaration
%\ioptwocol
%

\section{Introduction}

Virtually all intracellular processes, from metabolic reactions to gene expression, are subject to noise \cite{maheshri2007living}. Besides enabling complex phenomena like the coordinated expression of genes across large regulons \cite{stewart2012cellular} or the enaction of differentiation, stress response and adaptation strategies \cite{suel2007tunability}, noise is an inescapable source of phenotypic heterogeneity even in isogenic populations grown in the same environment \cite{rosenfeld2005gene,raj2008nature,raj2010variability,ackermann2015functional}. Quite remarkably, it has been known since the 1950's that a well-defined `average behaviour' can nevertheless  emerge, allowing to connect physiology and regulation by quantitative and reproducible  relationships at the population level \cite{monon2012growth,schaechter1958dependency,scott2011bacterial,hui2015quantitative,shahrezaei2015connecting}. The ongoing analysis of cell-to-cell variability, on the other hand, is shedding light on the structure and magnitude of biological noise at the phenotypic level \cite{gregor2007probing,dubuis2013positional,kiviet2014stochasticity,kellogg2015noise,avraham2015pathogen}. In particular, recent single-cell studies have quantified physiological growth-rate fluctuations across exponentially growing bacterial colonies in a variety of environments \cite{wang2010robust, ullman2013high, kennard2014individuality,iyer2014scaling}. Growth rate  distributions measured for different {\it E. coli} strains generically appear to be unimodal, with a well-defined mean and extended to remarkably fast rates. In addition, cross-correlations between doubling times and elongation rates have been detected, along with signatures of universality and weak dependence of results on cellular strains and/or growth media \cite{kennard2014individuality,iyer2014scaling,taheri2015cell}. These observations have allowed to settle long-standing cell biology issues like the mechanism behind cell size homeostasis in bacteria \cite{jun2015cell}. The origin of growth rate fluctuations, however, and specifically their relation to the underlying regulatory and metabolic activities of individual cells, is less clear.

Theoretical attempts at explaining observations have so far largely abstracted from the details of the biochemical and regulatory machinery of cells and focused on identifying general mechanisms behind cell-to-cell variability. Each of the key ingredients of available models --from the autocatalytic drive of cell growth \cite{iyer2014scaling,iyer2014universality}, to the coupling between growth and expression \cite{kiviet2014stochasticity}, to the need for an optimal scheduling of noisy cellular tasks \cite{pugatch2015greedy}-- likely captures relevant components of stochasticity in cell growth and division. Our proposal is simpler from a theory perspective but, being grounded on current biochemical knowledge (specifically, on genome-scale metabolic network reconstructions), allows in principle to directly connect empirical growth rate distributions  to their underlying regulatory scenarios. 

In essence, our argument is  based on the fact that the space of  feasible non-equilibrium steady states of {\it E. coli}'s metabolic network, which can be computed from the network's stoichiometry via constraint-based modeling, provides a map of the growth phenotypes that are in principle accessible to a population of non-interacting bacteria in a given environment.  A natural question to ask, then, is whether observed growth rate  distributions can be recovered  from specific, physically and/or biologically significant samplings of the phenotypic space. Following this route, we show that empirical data are reproduced by assuming an optimal (Maximum Entropy) trade-off between fast- and slow-growing phenotypes. We then introduce a minimal model for the evolution of a bacterial population in phenotypic space that generates this picture at stationarity, leading to a further characterization of the empirical scaling of growth rate distributions. Finally, we discuss the advantages, implications, potential applications, possible extensions and (several) limitations of our approach.

\section{Growth rate fluctuations in the space of metabolic phenotypes}

The standard {\it in silico} route to modeling cellular metabolic activity in a given medium relies on constraint-based models \cite{orth2010flux,lewis2012constraining,mccloskey2013basic,bordbar2014constraint,palsson2015systems}. In short, assuming that the intracellular reaction network that processes nutrients (e.g. glucose) to harvest free energy and synthetize macromolecular building blocks (e.g. amino acids) operates at a non-equilibrium steady state (NESS), feasible reaction flux vectors $\mathbf{v}=\{v_r\}$ ($r=1,\ldots,N$ indexing reactions) must satisfy the mass-balance conditions 
\begin{equation}\label{eq3}
\mathbf{S \cdot v}=\mathbf{0}~~,%~~~,~~~~~v_r \in [v_{r}^{{\rm min}},v_{r}^{{\rm max}}]
\end{equation}  
where $\mathbf{S}$ denotes the network's $M\times N$ stoichiometric matrix encoding for the input-output relationships underlied by each reaction (with $M$ the number of distinct compounds). Ranges of variability of the form $v_r \in [v_{r}^{{\rm min}},v_{r}^{{\rm max}}]$ must also be specified for each $r$ in order to account for thermodynamic irreversibility, kinetic bounds and other physiological or regulatory constraints. We further assume that $\mathbf{S}$ includes fluxes that determine exchanges of metabolites between the cell and the surrounding environment, thus defining the composition of the growth medium. In genome-scale models for specific organisms, $\mathbf{S}$ is  reconstructed from genomic data and, for a system such as {\it E. coli}, it represents a network with thousands of reactions and chemical species \cite{feist2009reconstruction}. Likewise, upper and lower bounds for fluxes are normally available based e.g. on prior biochemical knowledge. Therefore Eq. (\ref{eq3}) defines, for a given organism and a given medium, a high-dimensional  solution space $\mathcal{P}$ (a polytope) such that each vector $\mathbf{v}\in\mathcal{P}$ describes a feasible NESS of the network, or a `phenotype' for short. In turn, to each $\mathbf{v}\in\mathcal{P}$ a growth rate $\lambda\equiv\lambda(\mathbf{v})$ (measured in doublings per hour) can be associated through the corresponding biomass output \cite{feist2010biomass}, which for microbial systems is usually encoded in $\mathbf{S}$. A statistical sampling of $\mathcal{P}$ thus generates a distribution of growth rates relative to the specific environment defined in $\mathbf{S}$, to the regulatory constraints imposed by the bounds on fluxes, and to the chosen statistics. 

\begin{figure}
\centering\includegraphics[width=0.65\textwidth]{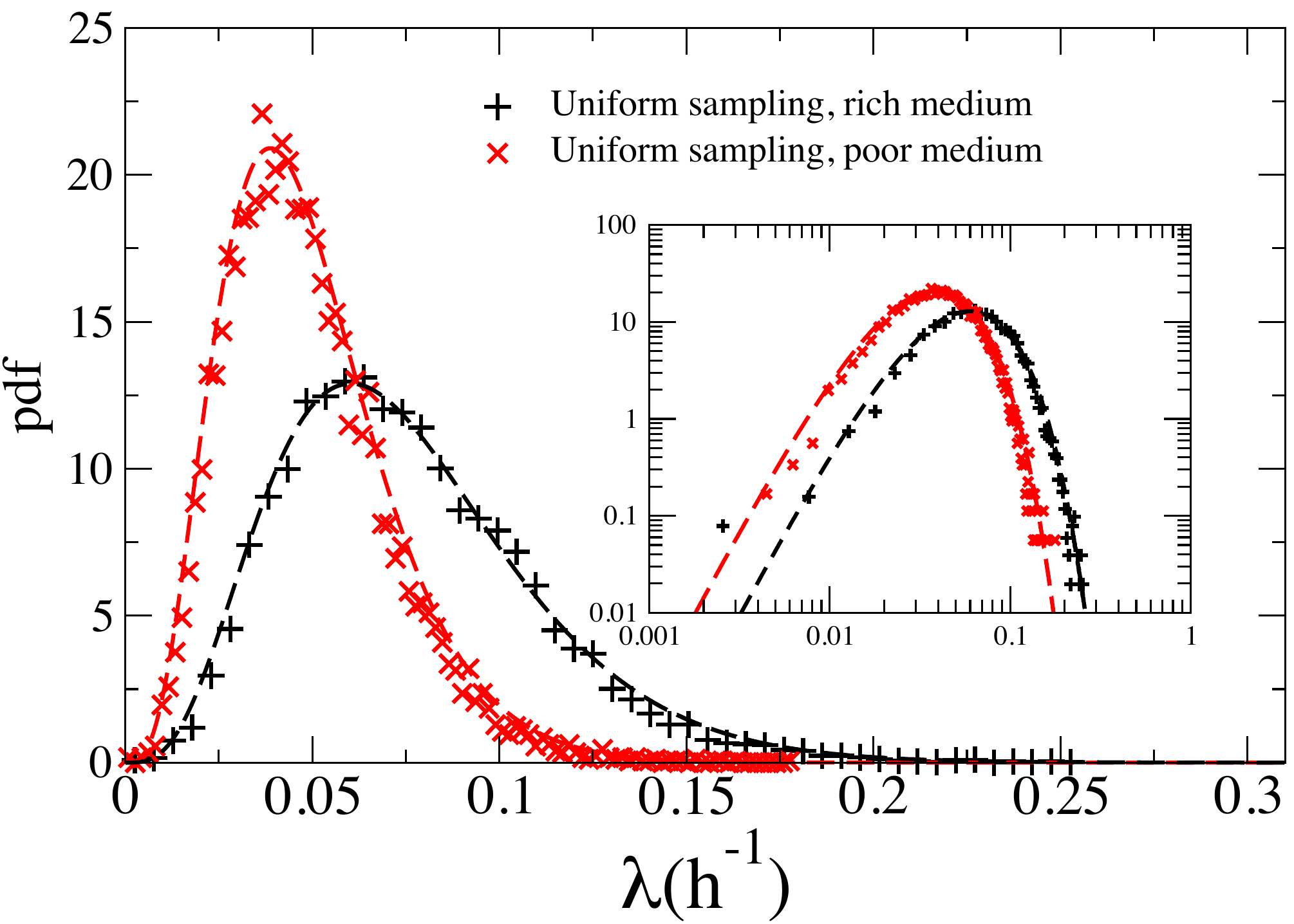}
\caption{\label{uno}Growth rate distributions for a uniform sampling of {\it E. coli}'s genome-scale metabolic network model iJR904, in a glucose-limited minimal medium as given in \cite{reed2003expanded}, with maximum glucose uptake $20$ mmol \gdw \hh and unlimited oxygen, with (rich medium, black markers) and without (poor medium, red markers) extra uptakes of arginine, lysine and phenylalanine, each with maximum uptake $10$ mmol \gdw \hh. Dashed lines show the best fit of the log of the curves. Inset: same on a log-log plot.}
\end{figure}

In what follows we shall focus on the metabolic network reconstruction iJR904 of the bacterium {\it E. coli} \cite{reed2003expanded}, which includes $1075$ intracellular  reactions among $761$ compounds together with details on the biomass composition and bounds on fluxes. The corresponding solution space of (\ref{eq3}) is a high-dimensional polytope (for instance, for a minimal glucose-limited medium one has ${\rm dim}(\mathcal{P})=233$) that can be sampled efficiently according to any prescribed probability distribution by a recently developed Hit-and-Run Monte Carlo method \cite{uniformell, capuani2015scirep}. For a uniform sampling of $\mathcal{P}$, the growth-rate distributions thus retrieved in a rich and a poor growth medium are shown in Fig. \ref{uno}.
Both curves are best fit by the formula
\begin{equation}
\label{eq:Q definition}
q(\lambda) \propto \lambda^b(\lambda_{\max}-\lambda)^a~~,
\end{equation} 
where $a>0$ and $b>0$ are constants, while $\lambda_{\max}$ is the maximum growth rate allowed by the constraints that define $\mathcal{P}$, which can be computed by linear programming as in the standard implementation of Flux-Balance-Analysis (FBA) for growth maximizing microbial systems \cite{orth2010flux}. Unsurprisingly, $\lambda_{\max}$ is found to be medium-dependent, with $\lambda_{\max}\simeq 1.9$ h$^{-1}$ and $\simeq 2.9$ h$^{-1}$ for the poor and rich glucose-limited media described in Fig. \ref{uno}, respectively. On the other hand, the fitting exponents $a$ and $b$ turn out to be medium-independent, thereby providing a robust characterization of the phenotypic space for carbon-limited growth. Specifically, we find $a\simeq 171$ and $b\simeq 3.6$ for {\it E. coli} iJR904\footnote{The fact that $a$ and $b$ are roughly constant suggests that, as long as glucose is the growth-limiting nutrient, the effective number of metabolic phenotypes leading to the same relative mean growth rate (i.e. to the same fraction of the maximum achievable mean growth rate) is roughly independent of the specific composition of the medium. In turn, growth in media in which other nutrients than glucose are limiting (like ammonia or oxygen) may lead to different values for the exponents $a$ and $b$. Likewise, more recent reconstructions are available (see e.g. \cite{feist2007genome,orth2011comprehensive}), where parts of the network, like exchange fluxes with the medium, are characterized in greater detail. Quantitative specifics (e.g. the values of $a$ and $b$) may differ from those found for iJR904. However we expect (\ref{eq:Q definition}) to remain valid.}. 

One sees that the overwhelming majority of feasible phenotypes corresponds to  slow-growing cells with growth rates about two orders of magnitude smaller than $\lambda_{\max}$. Empirical distributions in the same environments (see below) however concentrate around much faster rates. It is therefore tempting to think that observations may be explained in terms of a trade-off between dynamically favored, faster phenotypes and entropically favored, slower ones.

\section{Growth-entropy trade-off: MaxEnt growth rate distributions} 

The simplest way to represent such a trade-off within a probabilistic sampling scheme (i.e. without invoking a ``microscopic'', regulatory or population mechanism through which a growing bacterial colony may escape the entropic trap) is via the Maximum-Entropy (MaxEnt) framework\cite{bialek2012biophysics}. 
In brief, %fixing a mean growth rate induces a reduction of the volume of $\mathcal{P}$ and 
the MaxEnt distribution over a phenotypic space $\mathcal{P}$ is the one causing the smallest reduction in entropy of $\mathcal{P}$ at fixed mean growth rate $\langle\lambda\rangle$ and is, in this sense, the ``broadest'' and most unbiased distribution compatible with the constraint. A standard maximization of the entropy functional $S[f]=-\int f({\bf v})\log  f({\bf v}) d\bf{v}$ over distributions of flux vectors $f({\bf v})$ subject to a normalization constraint ($\int f({\bf v})d\bf{v}=1$) and at a fixed mean growth rate $\langle\lambda\rangle=\int\lambda({\bf v})f({\bf v})d{\bf v}$ yields
\begin{equation}\label{med}
f(\mathbf{v}) = \frac{e^{\beta \lambda(\bf{v})}}{Z(\beta)}~~~~~(\mathbf{v}\in\mathcal{P})~~,
\end{equation}
where $\beta>0$ is the Lagrange multiplier that constrains $\langle\lambda\rangle$ and
\begin{equation}
Z(\beta)=\int e^{\beta \lambda(\bf{v})}d{\bf v}~~.
\end{equation}
Correspondingly, the solution space entropy $S\equiv S(\beta)$ is reduced by a factor $I$ (measured in bits), given by
\begin{equation}
I \log 2 \equiv S(0)-S(\beta)
%= -\int \frac{1}{Z(0)}\log\frac{1}{Z(0)} d{\bf v}+\int \frac{e^{\beta \lambda}}{Z(\beta)} \log \frac{e^{\beta \lambda}}{Z(\beta)} d{\bf v}=\\
%-\log Z(0)+\int \frac{e^{\beta \lambda}}{Z(\beta)} \log \frac{e^{\beta \lambda}}{Z(\beta)} d{\bf v} = \\ \nonumber
= \beta \langle\lambda\rangle -\log \frac{Z(\beta)}{Z(0)}~~.
\end{equation}
On the other hand we have 
\begin{equation}
\frac{d }{d \beta}\log Z(\beta) = \langle\lambda\rangle~~,
\end{equation} 
where $\langle \lambda \rangle$ is a function of $\beta$ via $f$. Therefore, finally, 
\begin{equation}
 I \log 2=  \beta \langle \lambda \rangle-\int_0^\beta \langle \lambda \rangle d\beta' ~~.
%\log Z(\beta) = \int_0^\beta \langle \lambda \rangle d\beta' ~~,
\end{equation} 
%\red{ as follows. If we indicate with $Z(\beta)=\int e^{\beta \lambda} d{\bf v} $, we have for such entropy reduction
%\begin{eqnarray}
%I \log 2 = S(0)-S(\beta) = \\ \nonumber= -\log Z(0)+\int \frac{e^{\beta \lambda}}{Z(\beta)} \log \frac{e^{\beta \lambda}}{Z(\beta)} d{\bf v} = \\ \nonumber= \beta \langle\lambda\rangle -\log \frac{Z(\beta)}{Z(0)}
%\end{eqnarray}
%On the other hand we have 
%\begin{equation}
%\frac{d \log Z(\beta)}{d \beta} = \langle\lambda\rangle
%\end{equation} 
%where $\langle \lambda \rangle$ is a function of $\beta$ via $f$, from which finally we can calculate
%\begin{gather}
% I \log 2=  \beta \langle \lambda \rangle-\int_0^\beta \langle \lambda \rangle d\beta' ~~,
%\log Z(\beta) = \int_0^\beta \langle \lambda \rangle d\beta' ~~,
%\end{gather} 
%} 
The factor $\beta$ here mimics a ``selective pressure'' that allows to interpolate between entropy-dominated (low $\beta$) and growth-rate dominated  (high $\beta$) populations, the limit $\beta\to 0$ (resp. $\beta\to\infty$) corresponding to a uniform sampling of $\mathcal{P}$ (resp. a sampling concentrated on states with $\lambda=\lambda_{\max}$).

\begin{figure}
\centering\includegraphics[width=12cm]{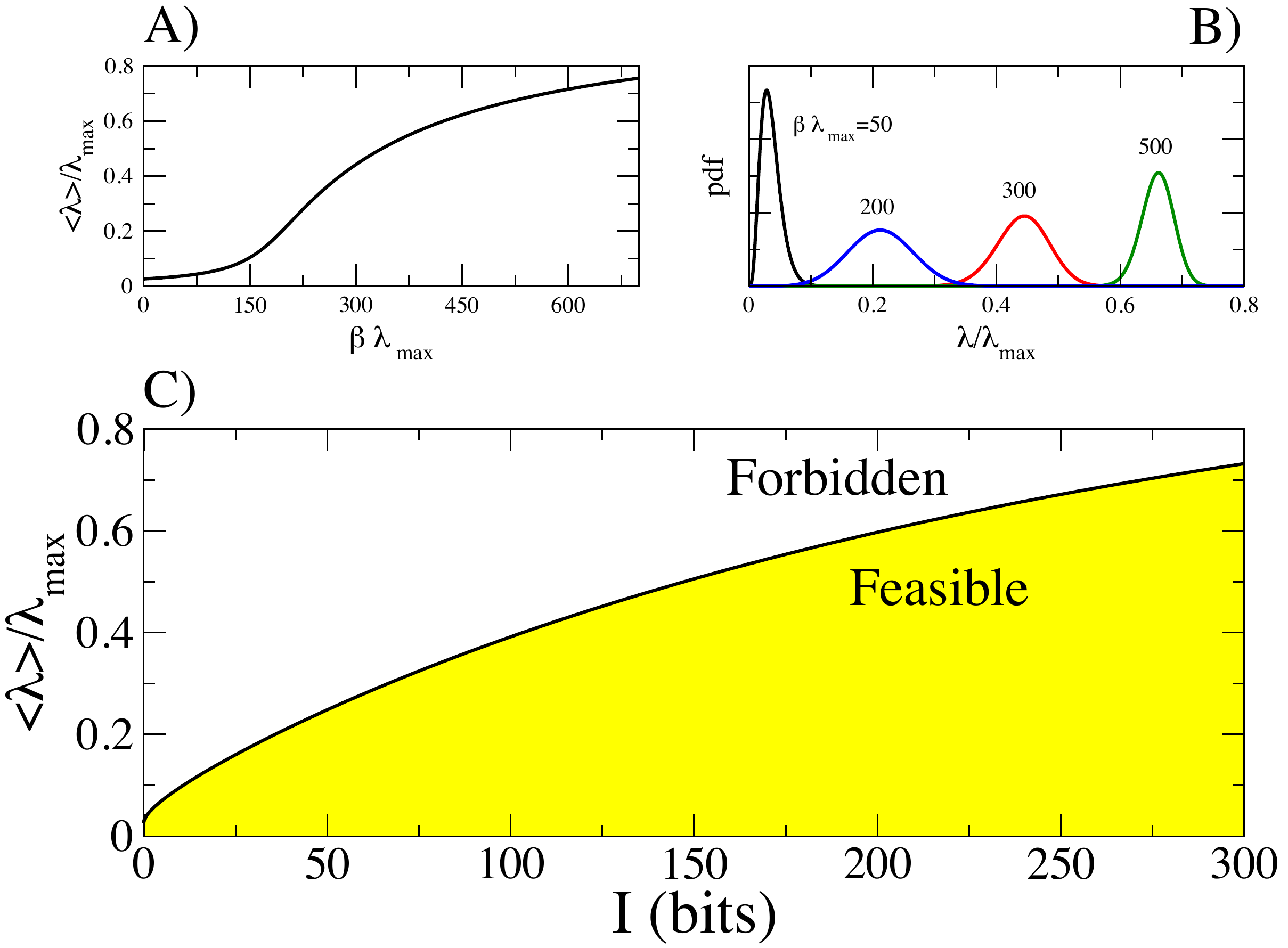}
\caption{\label{due}MaxEnt scenario for {\it E. coli} in a glucose-limited medium. A) Mean growth rate (in units of $\lambda_{\max}$) versus $\beta\lambda_{\max}$. B) MaxEnt growth rate distributions corresponding to selected values of $\beta\lambda_{\max}$. C) Mean growth rate (in units of $\lambda_{\max}$) versus the phenotypic space entropy reduction $I$ (measured in bits, so that $I=x$ implies that the effective volume of the phenotypic space is reduced by a factor $2^x$) in a glucose-limited medium. }
\end{figure}

\begin{figure}
\centering\includegraphics[width=0.45\textwidth]{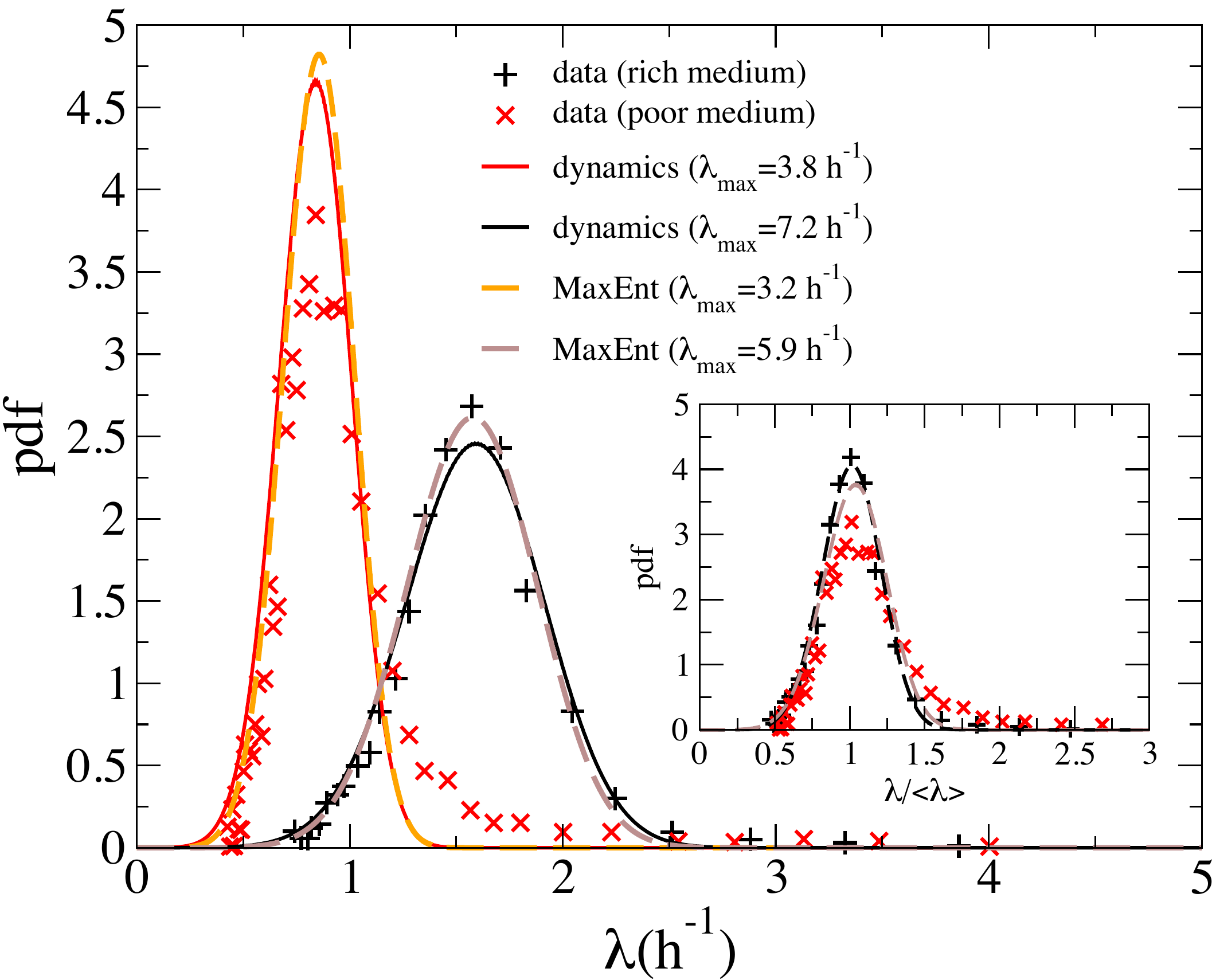}
\centering\includegraphics[width=0.5\textwidth]{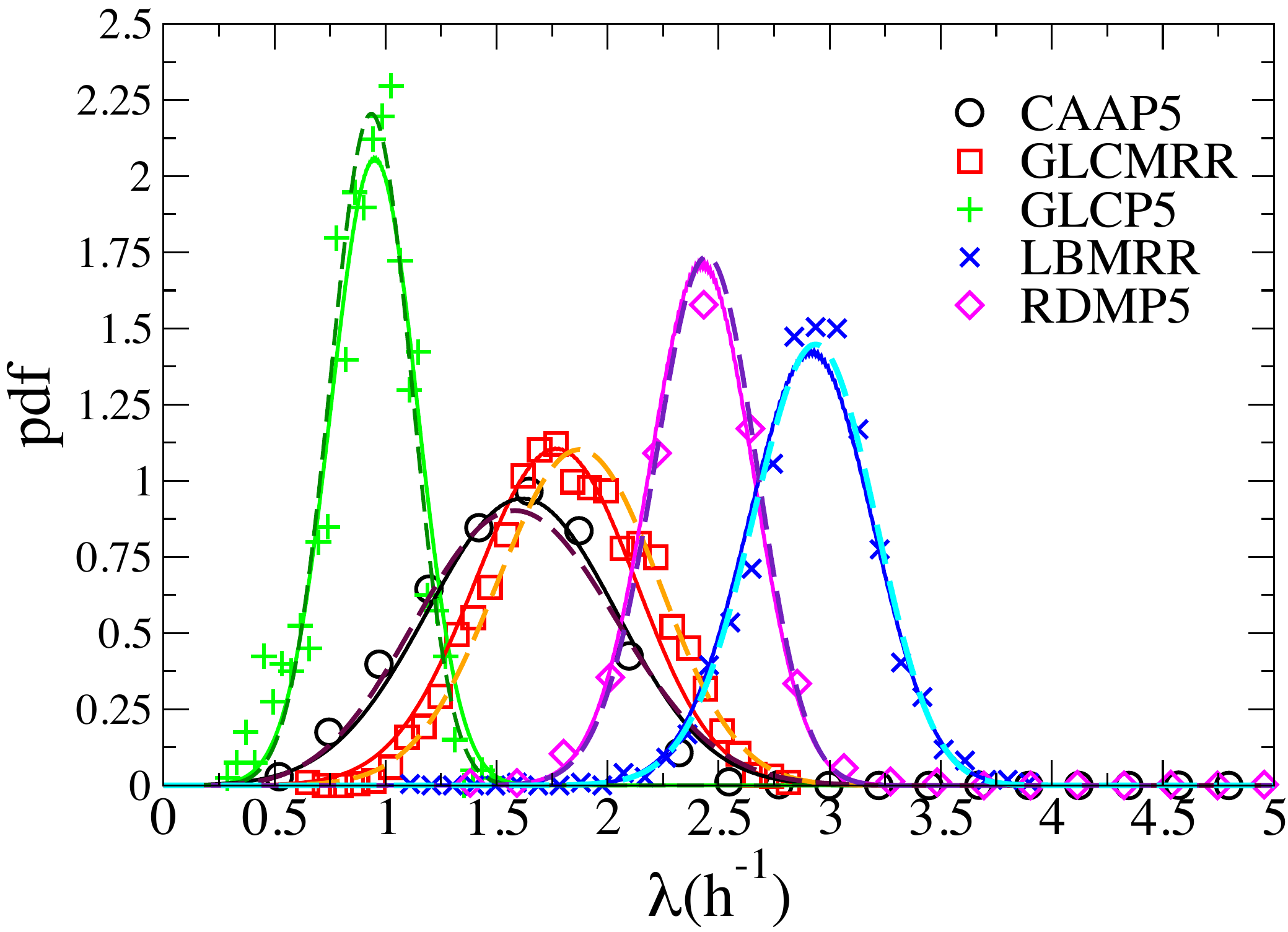}
\caption{\label{tre}Empirical growth rate distributions (markers) from \cite{ullman2013high} (left panel) and \cite{kennard2014individuality} (right panel), together with the best fitting MaxEnt distributions described in Section 3 (dashed lines in both panels) and the distributions derived from the dynamical model described in Section 4 (straight lines in both panels). Fitting parameters are summarized in Table \ref{tab}.}
\end{figure}

In this way, to each $\langle \lambda \rangle$ one can associate a minimal entropy reduction $I$, such that in order to achieve a mean growth rate $\langle \lambda \rangle$, the effective volume of the phenotypic space has to shrink at least by a factor of $2^I$. Vice-versa, to each $I$ one can associate a maximum achievable $\langle \lambda \rangle$, and achieving larger mean growth rates require larger values of $I$. This separates the $(I,\langle\lambda\rangle)$ plane in a `feasible' and a `forbidden' region (see \cite{bialek2012biophysics} for a related small-scale example). Results for a glucose-limited medium are shown in Fig. \ref{due} (a similar scenario holds in different media). 
%we show such $I(\langle \lambda \rangle)$ and in the insets $\langle \lambda \rangle (\beta)$ and growth rate distributions for particular values. 
%At odds with the dynamical case the average growth rate is smooth in the control parameter $\beta$.
The mean growth rate indeed increases with $\beta$ (panel A), while growth-rate distributions shift towards higher values as $\beta$ increases (panel B). The overall phase structure, shown in  panel C, quantifies the entropy reduction factor required to achieve a given $\langle\lambda\rangle$, with larger $\langle\lambda\rangle$'s requiring larger $I$'s. MaxEnt distributions lie on the line separating the feasible from the forbidden region in the $(I,\langle\lambda\rangle)$ plane, and can be fitted to empirical data via the parameters $\lambda_{\max}$ and $\beta$ (or $\beta\lambda_{\max}$). We have considered 7 data sets from different experiments \cite{ullman2013high, kennard2014individuality}. Fig. \ref{tre} shows the quality of the MaxEnt fits (dashed lines). Values of best fitting parameters are summarized in Table \ref{tab}.
\begin{table}
\begin{center}
\begin{tabular}{ | c  | c  |  c | c | c  |}  
\hline 
Data set & \multicolumn{2}{c |}{MaxEnt} & \multicolumn{2}{c |}{Dynamical} \\
 \cline{2-5}
 & $\lambda_{\max}$ [h$^{-1}$] &  $\beta \lambda_{\max} $ [adim.] & $\lambda_{\max}$ [h$^{-1}$] & $\sigma$ [adim.] \\ 
\hline 
\cite{ullman2013high} rich medium & $5.9$ & $220$ & $7.2$& $10^{-5}$   \\
\cite{ullman2013high} poor medium & $3.2$ &  $220$ & $3.8$ & $10^{-5}$ \\
\cite{kennard2014individuality} GLCP5  & $3.5$ &  $220$ & $4.3$ & $10^{-5}$ \\
\cite{kennard2014individuality} GLCMRR  &  $7$ &$220$ &$8$ &  $10^{-5}$ \\
\cite{kennard2014individuality} CAAP5  & $8.6$ &  $190$ & $9$ & $1.2\cdot 10^{-5}$  \\
\cite{kennard2014individuality} RDMP5  & $5.5$ &   $300$ & $6.4$ & $5 \cdot 10^{-6}$  \\
\cite{kennard2014individuality} LBMRR  & $6.6$ &   $300$ & $7.7$ & $5 \cdot 10^{-6}$ \\
\hline
\end{tabular}
\caption{\label{tab} Inferred maximum growth rates, level of optimization and rate of metabolic change for the experimental data \cite{ullman2013high, kennard2014individuality} fitted with the stationary distributions retrieved by the MaxEnt framework of Section 3 and from the dynamical model of Section 4. Growth rates are measured in h$^{-1}$, while $\sigma$ and $\beta\lambda_{\max}$ are adimensional.}
\end{center}
\end{table}
Notice that the same value of $\beta\lambda_{\max}$ (corresponding to the same `degree' of growth rate maximization in the MaxEnt framework) provides the best fit across four different data sets, while a 5th experiment (labeled CAAP5) appears to be very close to it. 
 %We do point out that the fit is pretty good for the body and right tail whereas the fit is not good for the left tail, especially for the case without amminoacids (data from left tail has been rejected in a previous modeling attempt). \red{This part needs to be rephrased once we find a way to tell the story of the left tail}
The fact that one can fit multiple experiments with the same values of $\beta\lambda_{\max}$ but different values of $\lambda_{\max}$ suggests that empirical distributions scale to the same ratio of the average to the maximum. Indeed, one has $\langle \lambda \rangle/\lambda_{\max} \simeq 0.28$ for each of these data sets. The collapse thus obtained for the data from \cite{ullman2013high} is shown in Fig. \ref{tre}, left panel. Similar collapses have been found in other experiments, including \cite{kennard2014individuality}.
The remaining two experiments (labeled RDMP5 and LBMRR) likewise appear to cluster at the same value of $\beta\lambda_{\max}$, in full agreement with the findings of \cite{kennard2014individuality}. Interestingly, from a physiological viewpoint, assuming empirical growth laws \cite{scott2011bacterial} can be extended to very fast rates, the fact that $\langle \lambda \rangle/\lambda_{\max}$ is the same for different bacterial populations suggests that they allocate to ribosomes a fixed share of the maximum lambda-dependent ribosomal proteome fraction.

\section{A minimal population-dynamical model}

The entropy-growth trade-off embedded in the MaxEnt scenario can be captured by a minimal population-dynamics model for the evolution of a (large) group of non-interacting bacteria. Denoting by $N(\lambda)$ the number of bacteria growing at rate $\lambda$, we assume that the population structure changes according to
%and change metabolism with rate $W(\lambda \to \lambda')$, i.e. we can write the equations
\begin{equation}
\dot{N}(\lambda)=\lambda N(\lambda) +\sum_{\lambda'} \left[W(\lambda' \to \lambda)N(\lambda')-W(\lambda \to \lambda')N(\lambda)\right] ~~,
\end{equation}
where $W(\lambda \to \lambda')$ stands for the transition rate from a phenotype with growth rate $\lambda$ to one with growth rate $\lambda'$. The first term on the right-hand side describes population increases due to replication events. The second term, instead, corresponds to (small) changes in growth rates due to metabolic re-arrangements that can be triggered e.g. by fluctuations in nutrient or enzyme availability or by variability induced in molecular levels at cell division. In terms of the population fractions $p(\lambda)=N(\lambda)/[\sum_{\lambda'}N(\lambda')]$, the above process takes the form 
\begin{equation}\label{pi}
\dot{p}(\lambda)=(\lambda-\langle \lambda \rangle) p(\lambda) +\sum_{\lambda'} \left[W(\lambda' \to \lambda)p({\lambda'})-W(\lambda \to \lambda')p({\lambda}) \right]~~,
\end{equation}
where $\langle \lambda \rangle =\int \lambda p(\lambda) d\lambda$. Phenotypic changes are assumed to occur so that the space $\mathcal{P}$ of viable phenotypes is explored in an unbiased way according to the detailed balance condition 
\begin{equation}
W(\lambda' \to \lambda)q(\lambda')=W(\lambda \to \lambda')q(\lambda)~~,
\end{equation}
where $q$ is the growth-rate distribution corresponding to a flat sampling of $\mathcal{P}$, given by (\ref{eq:Q definition}) in our case study. We also assume the existence of a fixed time-scale $\tau$, such that
\begin{equation}
\sum_{\lambda'} W(\lambda \to \lambda')=\frac{1}{\tau}~~.
\end{equation}
%in metabolism happen upon exploring the aforementioned distribution $P_e(\lambda)$ in unbiased way with some  timescale $\tau$:
%\begin{gather}
%\frac{w(\lambda' \to \lambda)}{w(\lambda \to \lambda')}=\frac{Q(\lambda)}{Q(\lambda')} \\
%\sum_{\lambda'} w(\lambda \to \lambda')=\frac{1}{\tau}~~.
%\end{gather}
%The rhs is a sum of replicator term and a master equation term.
If only transitions of the kind $\lambda\to\lambda\pm\delta$ with equal probability and with sufficiently small $\delta$ are allowed (implying that, generically, molecular fluctuations have a small impact on the growth rate), this scenario simply corresponds to a discrete random walk in the phenotypic space, so that $\tau$ represents the duration of a single time step and its inverse can be interpreted as the rate at which phenotypic changes occur. Under these conditions, the second term on the right-hand side of (\ref{pi}) can be expanded in a power series of $\delta$.
In the limit $\delta,\tau \to 0$, one obtains the non-linear Fokker-Planck equation
\begin{equation}\label{fp}
\dot{p}(\lambda) = (\lambda-\langle \lambda \rangle) p(\lambda) +D \left[\frac{\partial^2 p}{\partial\lambda^2} -\frac{\partial}{\partial\lambda} \left[p(\lambda) \frac{\partial}{\partial\lambda}(\log q(\lambda))\right]\right]~~,
\end{equation}
where $D=\delta^2/\tau$ represents the ``diffusion constant'' of the population in the phenotypic space. (Notice that it is the term proportional to $\langle \lambda \rangle$ that makes the above equation non-linear. Moreover, for $q=p$ Eq. (\ref{fp}) reduces to the replicator dynamics $\dot{p} = (\lambda-\langle \lambda \rangle) p$.) Once $q$ is fixed by (\ref{eq:Q definition}), with medium-independent values for the constants $a\simeq 171$ and $b\simeq 3.6$, the adjustable parameters left in the model are $D$ and $\lambda_{\max}$. For simplicity, one can re-scale $\lambda$ with $\lambda_{\max}$, which amounts re-scaling time as $t \to t\lambda_{\max}$. With this choice, $D$ also re-scales as $D \to D/\lambda_{\max}^3\equiv\sigma$. In this context, $\sigma$ corresponds to an effective rate of metabolic change.
%Upon passing to $\phi$ such that $p_\lambda=P_e \exp (\phi)$, the equation for $\phi$ is
%\begin{equation}
%\dot{\phi} = \lambda-\langle \lambda \rangle + D\left(  ( (\log P_e)'+\phi' )\phi'+\phi'' \right) 
%\end{equation}
%where the prime and the dot stand for derivation with respect to $\lambda$ and $t$, respectively. From now on, we re-scale $\lambda$ with respect to $\lambda_{max}$. This amounts at re-scaling times $t \to t\lambda_{max}$ and the only free parameter of the equation $D \to D/\lambda_{max}^3$. We divide the interval $[0,1]$ in $N$ equal parts of length $h=1/N$ and consider the discrete equation, upon considering discrete evaluation of the  derivatives:
%\begin{eqnarray}
%\phi_i' &=& \frac{-\phi_{i+2}+8\phi_{i+1}-8\phi_{i-1}+\phi_{i-2}}{12h} \quad 2 \leq i \leq N-3 \nonumber \\
%\phi_1' &=&\frac{\phi_2-\phi_0}{2h} \quad \phi_0' = \frac{\phi_1-\phi_0}{h} \nonumber \\
%\phi_{N-2}' &=&\frac{\phi_{N-1}-\phi_{N-3}}{2h} \quad \phi_{N-1}' =\frac{\phi_{N-1}-\phi_{N-2}}{h} 
%\end{eqnarray}
%and analogous formula for $\phi''$. The discrete time evolution equations read
%\begin{eqnarray}
%\phi_{i,t+dt} =\phi_{i,t}+f_i dt \nonumber \\
%f_i = i/N-\langle \lambda \rangle +D\left(  ( (\log P_e)_i'+\phi_i' )\phi_i'+\phi_i'' \right) 
%\end{eqnarray}
%From now on we consider the uniform distribution retrieved from the aforementioned genome scale metabolic network model of E Coli and we take $N=10^3$ and $dt=10^{-4}$. 

\begin{figure}
\centering\includegraphics[width=10.25cm]{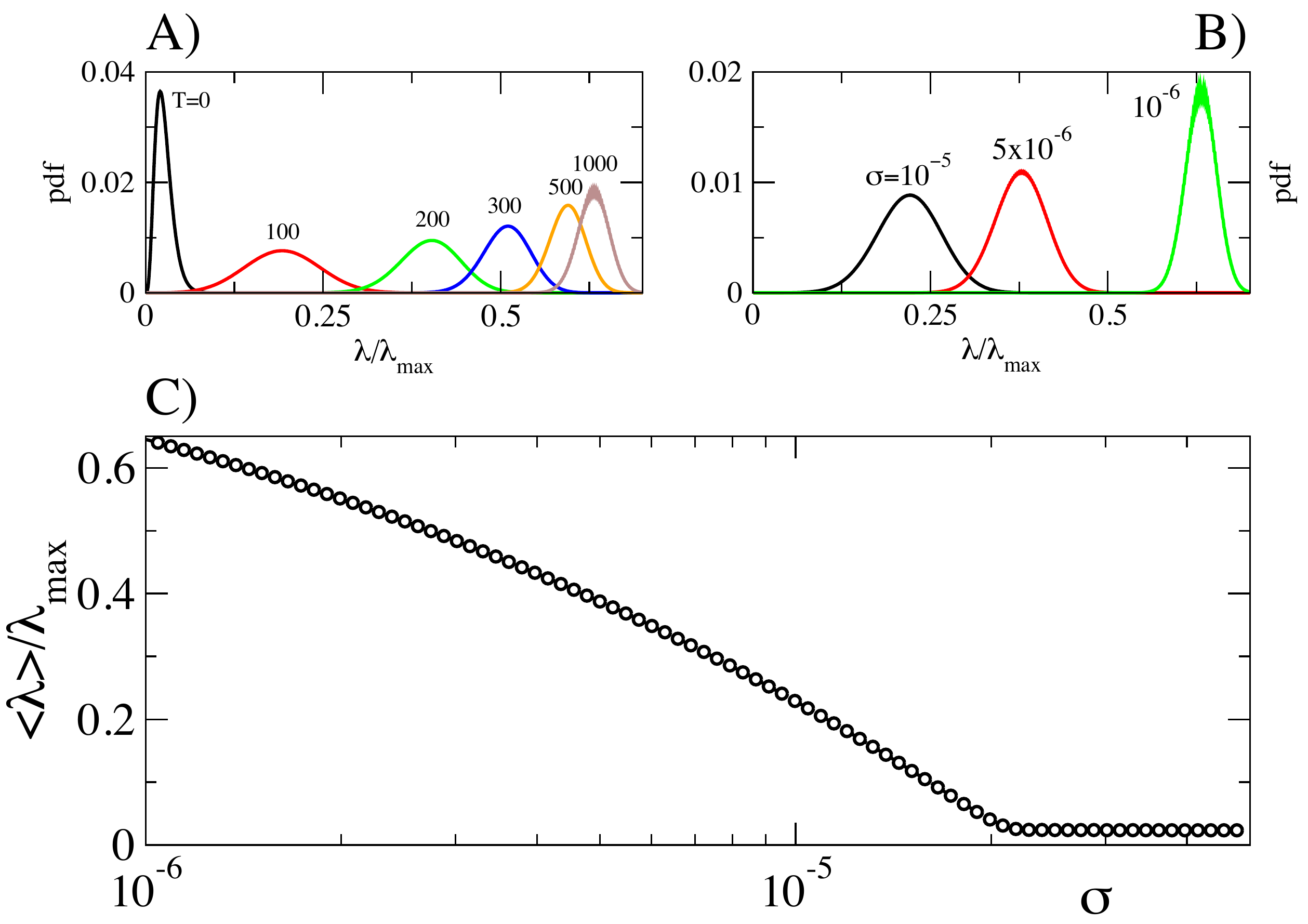}
\caption{\label{quat}Results from the minimal population dynamical model.  A) time-evolution of $p(\lambda)$ for $\sigma=10^{-6}$ ($T=$ number of time steps in units of $\lambda_{\max}^{-1}$), B) stationary growth rate distributions obtained for different values of $\sigma$, and C) stationary mean growth rate $\langle \lambda \rangle$ as a function of $\sigma\equiv D/\lambda_{\max}^3$.}
\end{figure}

Numerical solution of (\ref{fp}) leads to the scenario described in Fig. \ref{quat}. Panel A displays the time-evolution of $p(\lambda)$ obtained for $\sigma= 10^{-6}$. One sees that a stationary distribution is attained after roughly $10^3$ time steps (in units of $\lambda_{\max}^{-1}$). 
%\begin{figure}[h!!!]
%\includegraphics[width=0.45\textwidth]{evolution}
%\caption{Evolution of the growth rate distribution during the growth dynamics after $T$ time steps starting from an uniform prior in  the metabolic space  for $D/\lambda_{max}^3=10^{-6}$. \red{not sure I'd leave this; maybe unify with the next Fig?}}
%\end{figure}
The form of the stationary distribution on the other hand depends on $\sigma$ as shown in panel B. As expected, smaller $\sigma$'s (i.e. smaller diffusion constants or smaller rates of phenotypic change) allow the population to settle at higher growth rates. Therefore, $\sigma$ here plays a role analogous to $1/\beta$ in the MaxEnt scenario. The mean growth rate $\langle \lambda \rangle$ indeed decreases as $\sigma$ increases and its derivative appears to change discontinuously at $\sigma\simeq 2~10^{-5}$ (Panel C). This behaviour should be compared with the MaxEnt scenario, shown in Fig. \ref{due}B and C.
%We will fit experimental data with the distributions retrieved by these models in the following section.
%\section*{Maximum entropy distributions}
%\section*{Comparison with experimental data and scaling}
%The empirical distribution of generation times, defined as $\tau =\log 2/\lambda$ \red{comment}, from ELF referring to batch cultures in a glucose limited minimal medium with and without amminoacids. 
Stationary growth rate distributions obtained by this model can  be fitted against empirical ones, the fitting parameters now being  $\sigma$ and $\lambda_{\max}$. The quality of the fits is shown in Fig. \ref{tre} (continuous lines), whereas the values of best fitting parameters are summarized in Table \ref{tab}. Again, $\sigma\simeq 10^{-5}$ for five different data sets, while $\sigma\simeq 5\cdot 10^{-6}$ for the remaining two, in full agreement with the MaxEnt scenario and with previous analyses of the experimental data. Values of $\lambda_{\max}$ instead appear to be systematically larger than those obtained within the MaxEnt scenario, albeit similar. These differences however can account, at least qualitatively, for part of the discrepancies one sees between the MaxEnt and population-based distributions.

%\section*{Conclusions}

\section{Discussion}

%Novel single cell techniques
Many recently developed techniques provide access to the growth physiology of a bacterial population at single cell resolution, allowing to enrich the picture underlied by average `growth laws' \cite{scott2011bacterial} by characterizing fluctuations in some key observables. In this work we have addressed the origin of empirical growth rate distributions in {\it E. coli}, with the aim of connecting them to the underlying cellular activity by `minimal' physically significant assumptions. Adopting the standard constraint-based modeling framework for cell metabolism, and employing an efficient method to sample the solution space of {\it E. coli}'s iJR904 metabolic network reconstruction, we have applied statistical physics reasoning to obtain the following results.
\begin{enumerate}
\item The bare growth rate distribution obtained through a flat sampling of the solution space is dominated by slow growing phenotypes, i.e. with doubling times much larger than the smallest compatible with constraints. This is not surprising in view of the fact that growth is loosely constrained by standard physiological bounds on fluxes.
\item MaxEnt distributions at fixed mean growth rates, corresponding to a minimal ``static'' assumption on the way in which a population of non-interacting bacteria organizes in the phenotypic space, describe empirical data for different bacterial strains and environments by tuning two parameters, namely the maximum achievable growth rate $\lambda_{\max}$ and the `inverse temperature' $\beta$ that fixes the mean growth rate. Empirical data can thus be seen as an optimal trade-off between the few, dynamically favored fast growing phenotypes and the many, entropically favored slow growing ones.
\item This trade-off can be captured within a minimal dynamical  model, leading to a qualitatively similar scenario in which the role of a novel parameter (the rate of phenotypic change) is highlighted. In the light of both the static and the dynamical view, we have offered a novel interpretation of the scaling properties found in empirical distributions.
\end{enumerate}

\begin{figure}
\centering\includegraphics[width=0.65\textwidth]{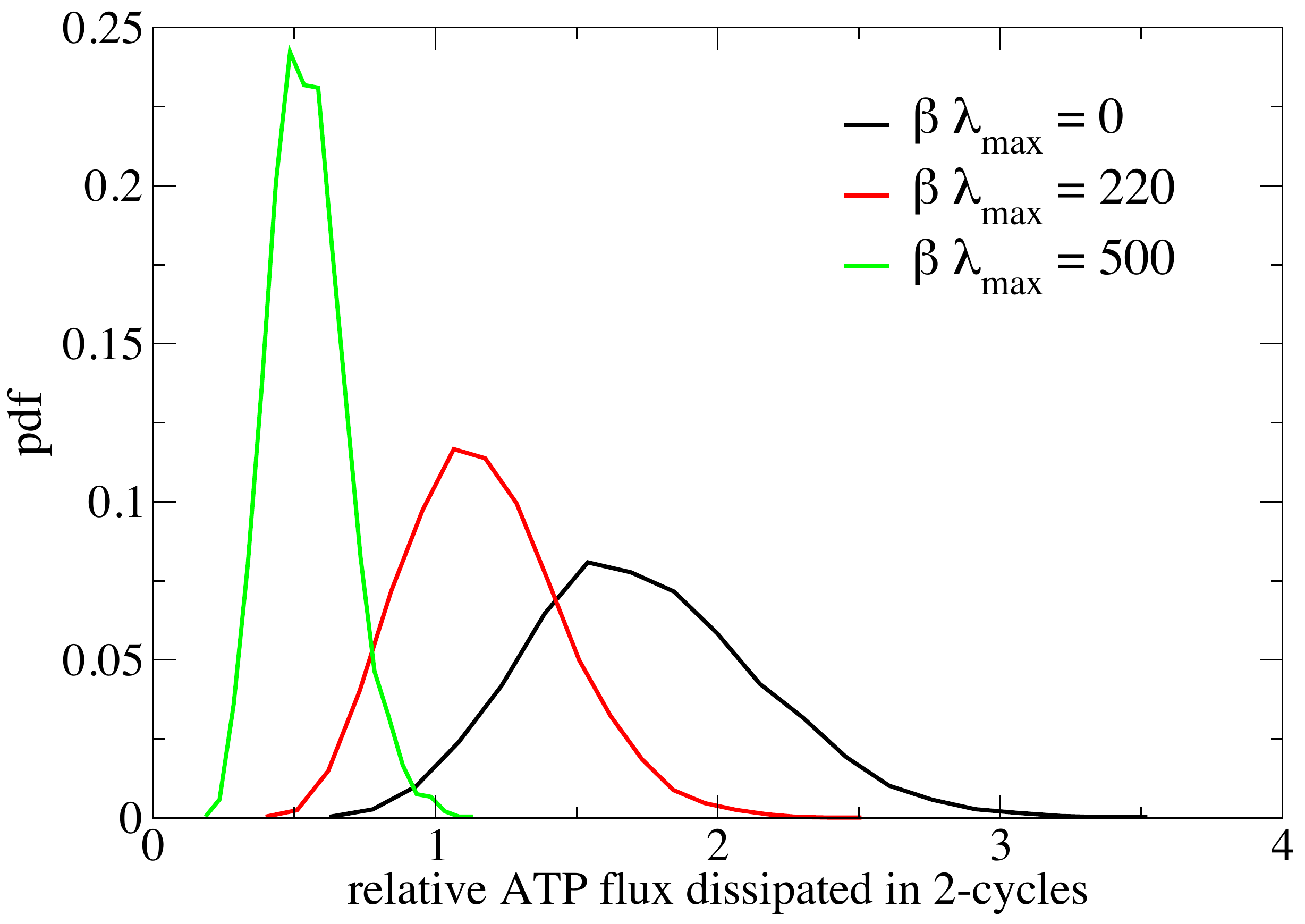}
\caption{\label{futile}Distribution of the ATP flux dissipated in futile cycles of length 2 (in units of the maximum glucose uptake) for {\it E. coli}'s metabolism in minimal glucose-limited medium obtained by MaxEnt sampling at different values of $\beta\lambda_{\max}$ (larger $\beta\lambda_{\max}$ implying larger mean growth rate). In the limit $\beta\to\infty$, corresponding to the solution(s) of FBA, ATP dissipation vanishes.}
\end{figure}

The approach described here presents in our view two merits. The first one lies in its simplicity: both of the views we discuss recapitulate the empirical evidence on growth rate distributions from minimal assumptions. In addition, being fully rooted in available network reconstructions, they allow in principle to associate growth rate distributions to (hopefully) specific states of a cell's metabolism, since cells must actively repress inefficient metabolic pathways in order to drive the population toward faster growing phenotypes. For instance, a deeper look at how solutions change as the mean growth rate is pushed to higher values reveals that increases in fitness $\langle\lambda\rangle$ are tightly linked to a decrease of redundancy in flux patterns, more precisely via reduction of flux through the so-called futile cycles\footnote{A futile cycle occurs when, in a solution, two distinct reactions or pathways are active in opposite directions, leading to no other net effect than the consumption (hydrolysis) of ATP.} (see Fig.  \ref{futile}).
In other terms, as $\beta$ increases, solutions tend to become more efficient by lowering energy dissipation. 

\begin{figure}
\centering\includegraphics[width=10.25cm]{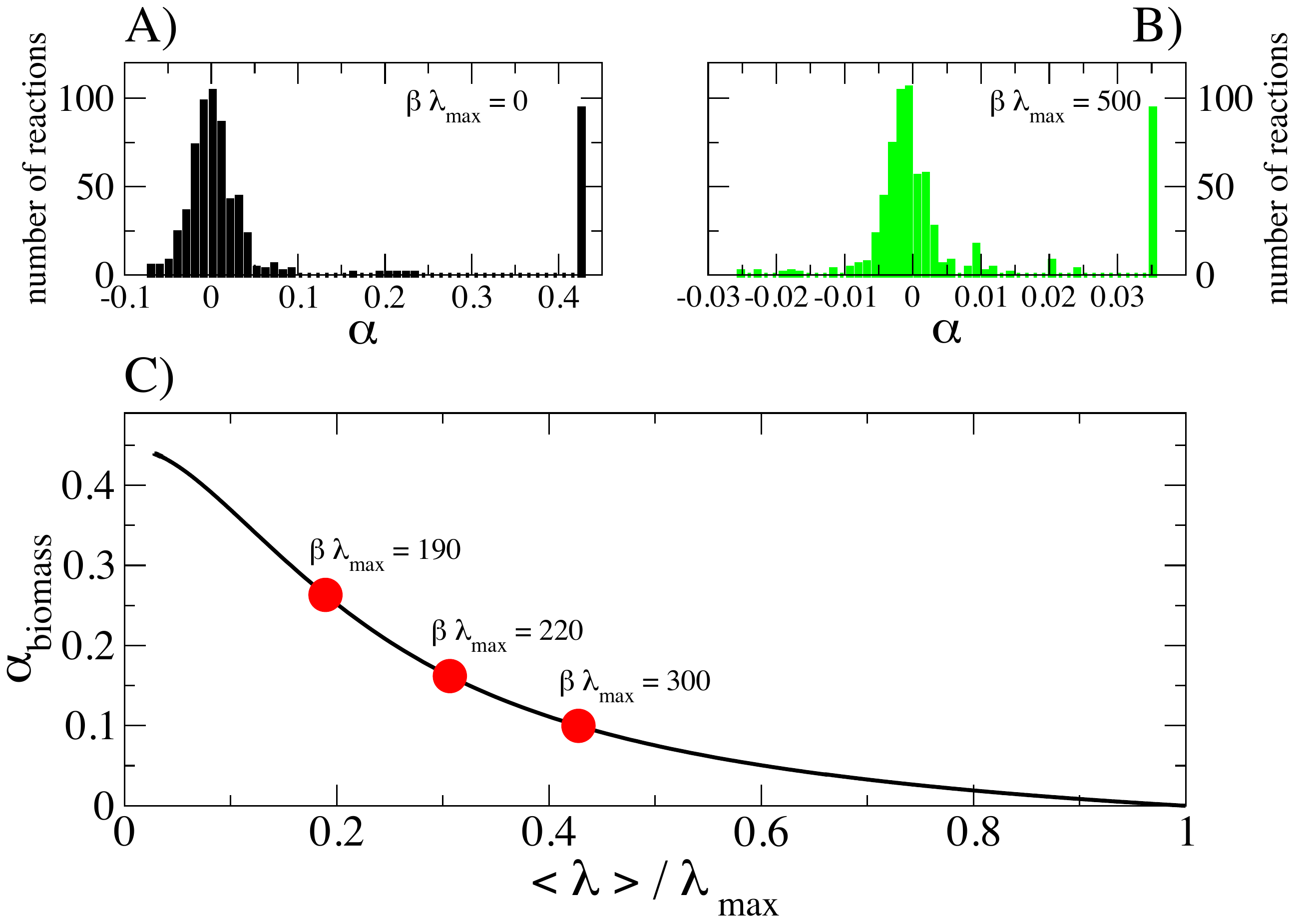}
\caption{\label{alpha}Histogram of optimization potentials of individual {\it E. coli} reactions derived from a uniform sampling (A) and from a MaxEnt sampling with $\beta\lambda_{\max}=500$ (B). (C) Optimization potential for the biomass reaction, equal by definition to the relative fluctuations of its flux (which can be computed directly from (\ref{med})), as a function of the fitness $\langle\lambda\rangle$ re-scaled by $\lambda_{\max}$. Red markers give the optimization potentials of the bacterial populations described in Table \ref{tab} and characterized by different inferred  values of $\beta\lambda_{\max}$.}
\end{figure}

Given the complexity of the underlying biochemical machinery, it is somewhat striking that the MaxEnt rule appears by itself to be able to account for such mechanisms. Previous models, which mainly aim at isolating key general features from the complicated growth physiology, certainly capture important aspects of the problem that our approaches are  unable to bring to light \cite{kiviet2014stochasticity,iyer2014scaling,iyer2014universality,pugatch2015greedy}. On the flipside, though, they don't allow for an immediate connection with known regulatory and/or biochemical elements. In this respect, it is  interesting to study an ``optimization potential'' associated to each reaction $i$. Inspired by the notion of selection strength \cite{evol}, we define it in terms of the relative change of fitness obtained when the net flux through $i$ takes the value $v_i^\star=\langle v_i\rangle+\sigma_{v_i}$, where $\langle v_i\rangle$ and $\sigma_{v_i}$ are the mean and variance of the distribution of $v_i$, i.e.
\begin{equation}
\alpha_i=\frac{\langle \lambda|v_i^\star\rangle}{\langle \lambda\rangle}-1~~,
\end{equation}
where $\langle \lambda|v_i^\star\rangle$ is the mean growth rate conditional on $|v_i|=v_i^\star$. Values of $\alpha_i$ close to zero suggest that reaction $i$ does not contribute significantly to the fitness, so that the conditional average $\langle \lambda|v_i^\star\rangle\simeq \langle \lambda\rangle$. On the other hand, large positive values of $\alpha_i$ point to a key role of reaction $i$ in determining $\langle \lambda\rangle$. Fig. \ref{alpha}A-B shows that, as $\beta$ increases, the distribution of the $\alpha_i$'s becomes more and more concentrated around 0, with a significant peak at positive $\alpha$'s corresponding to roughly 90 reactions carrying a strong correlation with the biomass flux (the position of the peak indicating the relative fitness gain obtained by perturbing those reactions as described above).
The optimization potential for the biomass reaction itself, which equals its relative fluctuations (i.e. $\alpha_{{\rm biomass}}=\sigma_{v_{{\rm biomass}}}/\langle\lambda\rangle$), expectedly gets smaller as $\langle \lambda\rangle$ increases (see Fig. \ref{alpha}C). Bacterial populations grown in richer media turn out to have, in this sense, a smaller optmization potential than populations grown in poorer conditions.

A more detailed analysis and a more careful calibration of prior biochemical information (work along these lines is in progress) is needed to break down the contribution of different individual pathways to the above scenario. However, even without going into biochemical details, these results present two straightforward insights of potential relevance for biology. The first is shown in the ``phase structure'' of Fig. \ref{due}C. With our choices for the units, if one could view metabolism as a ``binary network'' in which reactions can be either on or off, the $x$-axis would roughly indicate the number of variables (fluxes) that need to be constrained in order to achieve the desired entropy reduction and mean growth rate. The fact that, in the continuous flux model of cellular metabolism, $I$ can exceed the dimension of the phenotypic space suggests that simple regulatory strategies that either silence or activate reactions in order to achieve a certain objective are unlikely to suffice: a finer degree of flux regulation is required. For a second insight, we notice that the values of $\lambda_{\max}$ obtained by fitting empirical data represent {\it bona fide} predictions for the maximum growth rates achievable by a {\it single} {\it E. coli} bacterium (as opposed to the mean growth rates of a population of cells) in different media. To our knowledge, no other method of analysis provides a similar information.

Among the main limitations of our approach is the fact that, by  relying on a fixed phenotypic space, we are implicitly not accounting for cell-cell interactions and cell-to-cell fluctuations in nutrient availability. In this sense, we are  considering a low density bacterial population in a well mixed growth medium. Secondly, the dynamical model predicts trivial mother-daughter correlations, at odds with observations that suggests a much richer picture (see e.g. \cite{ullman2013high}, Supporting Material). That framework should therefore be seen as a zero-order description of the actual population dynamics, and will therefore need to be substantially enriched in order to get closer to biological reality (and, possibly, closer to the MaxEnt scenario).

\section*{Acknowledgments}

We are indebted with Pietro Cicuta, Marco Cosentino Lagomarsino and Matteo Osella for sharing data and for very useful comments and criticism. The research leading to these results has received funding from the People Programme (Marie Curie Actions) of the EU Seventh Framework Programme (2007-2013) under REA grant agreement n. 291734, and from the Marie Curie Action ITN NETADIS, grant agreement n. 290038.

\section*{References}

\bibliographystyle{unsrt}
\bibliography{notabib}

\end{document}